# *h*-index Research in Scientometrics: A Summary

Lutz Bornmann

Division for Science and Innovation Studies

Administrative Headquarters of the Max Planck Society

Hofgartenstr. 8,

80539 Munich, Germany.

Email: bornmann@gv.mpg.de

Abstract:

A Letter to the Editor shortly summing up ten or so years of research into the *h*-index.

Dear Sir,

Significant and important developments in a research area are usually expected to come from well-known experts in the field. However, the development of the *h*-index did not conform to this expectation: a physicist, Jorge Hirsch, who had never before published a paper on bibliometrics proposed an indicator – the *h*-index (Hirsch, 2005) – which resulted in a new research front in bibliometrics. What can we conclude from the scientific investigations into this indicator which have taken place over the last ten years? As Figure 1 shows, the popularity of the *h*-index in bibliometrics continues unabated. It lists the number of papers that have been published in the *Journal of Informetrics*, in the *Journal of the American Society for Information Science and Technology* (now *Journal of the Association for Information Science and Technology*), in *Scientometrics* and in *Research Evaluation* and which have cited the paper on the h index.

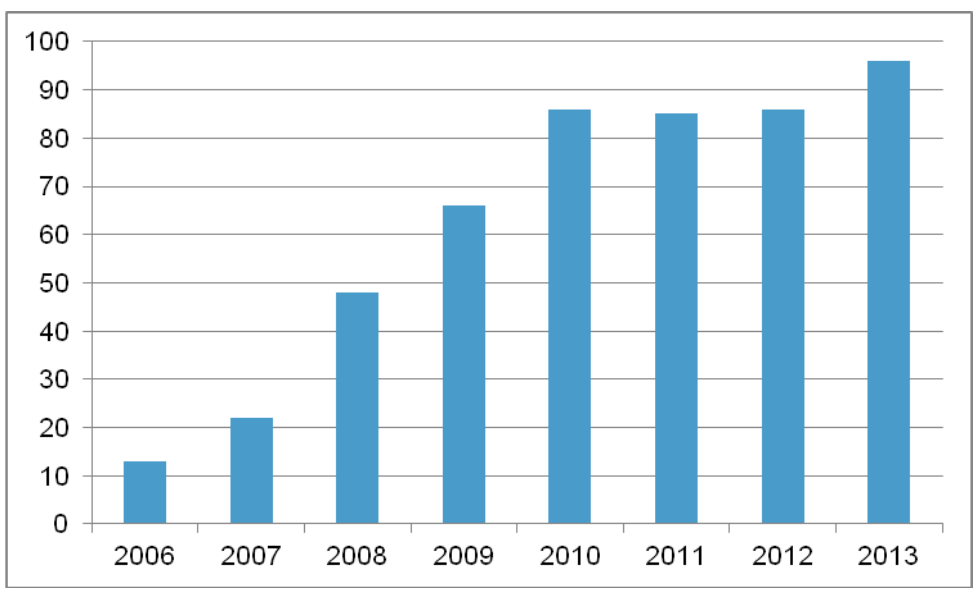

Figure 1. Number of papers in selected journals which have cited the paper by Hirsch (2005)

Even though it was undoubtedly useful to develop an indicator which linked productivity and citation impact, it is still questionable as to whether the link created by Hirsch (2005) was meaningful. "For instance, the *h*-index could equally well have been defined as follows: A scientist has an *h*-index of h if h of his publications each have at least 2h citations and his remaining publications each have fewer than 2(h+1) citations. Or the following definition could have been proposed: A scientist has an *h*-index of h if h of his publications each have at least h/2 citations and his remaining publications each have fewer than (h+1)/2 citations" (Waltman & van Eck, 2012, p. 408). Identifying publications with a certain citation impact with the h index is arbitrary because it has no inherent basis. For decades, bibliometrics has developed generally accepted methods which identify the significant publications in a publication set (such as those publications, for example, which are among the 10% most cited publications in their subject area and publication year). The benchmark for significant publications should therefore not result from the publication set of a scientist, as is the case with the h index; the <u>same</u> benchmark should be used for the publications of <u>all</u> scientists to evaluate whether a publication should be designated significant or not.

As the *h*-index does not take account of any normalization of citation impact regarding the publication year and the subject area, it cannot be used to compare individuals who have published in different subject areas and publication years. Furthermore, the indicator cannot be used to compare people with different academic ages as the expected values for publications and citations are different depending on the age. These two limitations inhibit the use of the indicator in research evaluation almost entirely, as the scientists as a rule have not published in similar subject areas and publication years, nor have a similar academic age. Better indicators have been proposed for comparisons across subject areas, publications years and ages (Bornmann & Marx, 2014).

Research into the *h*-index has so far yielded around 50 proposed variants which allegedly do not exhibit certain of its disadvantages (Bornmann, Mutz, Hug, & Daniel, 2011). It is interesting that none of the variants has achieved major significance in research evaluation. It is inevitable that bibliometricians question the meaningfulness of this research, which has looked at the development, testing and comparisons of the variants. Only the *h*-index itself has been integrated in two important multidisciplinary databases – Web of Science (Thomson Reuters) and Scopus (Elsevier) – and is available to users along with other standard indicators.

Nowadays, the *h*-index is used primarily by those who are not bibliometrics specialists. It appears therefore that a similar fate to that of the Journal Impact Factor awaits the *h*-index (Bornmann, Marx, Gasparyan, & Kitas, 2012). This is another indicator which is also very popular with non-professional bibliometrics. Discussion of both the Journal Impact Factor and the *h*-index in bibliometrics has been chiefly critically. There are probably few disciplines other than bibliometrics in which a response outside of the discipline has been so little shaped by the standards operating within the discipline. It remains to be hoped this will change in the future.